\documentclass[british]{article}
\usepackage[OT1]{fontenc}
\usepackage{textcomp}
\usepackage[latin9]{inputenc}
\usepackage[a4paper]{geometry}
\usepackage{color}
\usepackage{babel}
\usepackage{float}
\usepackage{url}
\usepackage{amstext}
\usepackage{graphicx}
\usepackage{setspace}
\usepackage[authoryear]{natbib}
\usepackage[pdfusetitle,
 bookmarks=true,bookmarksnumbered=false,bookmarksopen=false,
 breaklinks=false,pdfborder={0 0 1},backref=false,colorlinks=false]
 {hyperref}
\hypersetup{
 colorlinks=true,citecolor=magenta,linkcolor=blue}

\makeatletter
\newcommand{\lyxaddress}[1]{
	\par {\raggedright #1
	\vspace{1.4em}
	\noindent\par}
}

\date{}

\makeatother

\begin{document}
\title{\noindent gpuGEM: GPU-based linear optimisation of genome-scale metabolic
\textcolor{black}{models}}
\author{Farid Zare$^{1,2,3\dagger}$, Timothy Huls\textcolor{black}{ho}f$^{1,2\dagger}$,
Ines Thiele$^{1,2,4,5,6}$, Ronan M.T. Fleming$^{1,2,3*}$}
\maketitle
\begin{singlespace}

\lyxaddress{\textsuperscript{1}School of Medicine, University of Galway, University
Rd, Galway, Ireland; \textsuperscript{2}Digital Metabolic Twin Centre,
University of Galway, Galway, Ireland; \textsuperscript{3}Institute
for Clinical Trials, University of Galway, Galway, Ireland; $^{4}$
School of Microbiology, University of Galway, Galway, Ireland; $^{5}$
Ryan Institute, University of Galway, Galway, Ireland; $^{6}$ APC
Microbiome Ireland, Cork, Ireland}
\end{singlespace}

$\dagger$ These authors contributed equally to this work.
\begin{abstract}
\textbf{Summary:} \textcolor{black}{Genome-scale metabolic models
increasingly challenge CPU linear-programming solvers in repeated
optimisation. We present gpuGEM, an open-source Python library for
GPU-based optimisation using NVIDIA cuOpt, with size-adaptive settings
validated against HiGHS and Gurobi. Across ten models spanning four
orders of magnitude, gpuGEM achieved 1.1$\times$ to 94$\times$ speed-ups
over the faster HiGHS method (dual simplex or interior point), and
outperformed Gurobi on three models (1.6\texttimes{} to 2.5\texttimes ).
In candidate-gene prioritisation, gpuGEM reduced end-to-end query
time from 41\,s to 21\,s.}

\textbf{Availability and Implementation:} gpuGEM is released under
the MIT licence at \url{https://github.com/Digital-Metabolic-Twin-Centre/gpuGEM}
and on PyPI (pip install gpugem); release v0.1.3 is archived at \url{https://zenodo.org/records/23422813}.
It requires an NVIDIA GPU of compute capability 7.0 or newer (Volta
onward), Python 3.11 or later, CUDA 12 and cuOpt 26.6 or later.

\textbf{Contact:} ronan.mt.fleming@gmail.com

\textbf{Supplementary information:} Supplementary data are available
as an ancillary file on arXiv.
\end{abstract}

\section{Introduction}

Genome-scale metabolic models (GEMs) link genes, reactions, metabolites,
and constraints from cells to whole organisms \citep{gu_current_2019,thiele_protocol_2010a}.
Flux balance analysis (FBA) predicts feasible fluxes by solving a
linear programme under steady-state mass balance \citep{orth_what_2010}.
Gene-deletion screens, flux variability analysis, model personalisation,
and cohort studies require hundreds to thousands of related solves,
making solver runtime an important determinant of computational feasibility.
On our benchmark workstation, a single FBA solve of the Harvey whole-body
model (81,094 reactions, 103,600 coupling constraints) takes 33\,s
with HiGHS, making repeated optimisation costly.

Standard FBA relies on CPU linear programming solvers. As genome-scale
modelling extends into metabolic digital twins and human health applications,
reproducibility and auditability become increasingly important, strengthening
the case for an openly available solver stack. HiGHS\textcolor{black}{{}
\citep{huangfu_parallelizing_2018} provides an open-source CPU alternative.
Meanwhile, advances in artificial intelligence have accelerated investment
and innovation in GPU computing. First-order optimisation methods
avoid repeated matrix factorisations and rely heavily on sparse matrix-vector
operations, making them well suited to GPU execution. Extending genome-scale
metabolic modelling to GPUs therefore offers an opportunity to benefit
from continued advances in GPU hardware and software. NVIDIA cuOpt
provides GPU implementations of large-scale linear programming methods,
including PDLP, a restarted primal-dual hybrid gradient method \citep{applegate_practical_2021,lu_cupdlpjl_2023,lu_cupdlpc_2023}.
The field has moved quickly since: cuPDLPx reports further gains over
cuPDLP-C \citep{lu_cupdlpx_2025,lu_overview_2025}, and first-order
GPU modes now ship in HiGHS and in Gurobi 13 \citep{highs_pdlp_2025,gurobi_optimizer_2025}.
The contribution here is therefore not the solver itself but its integration
into a constraint-based modelling workflow, together with the size-adaptive
settings and validation that make it usable on genome-scale and whole-body
models. Prior GPU work in constraint-based modelling has addressed
flux sampling \citep{benguebila_achr.cu_2019} and parallel pathway
analysis \citep{arabzadeh_system_2020}, rather than GPU solving of
GEM-derived linear programmes. To our knowledge, no previous study
has systematically evaluated and validated GPU-based linear programming
for GEMs at whole-body and personalised microbiome scale. The central
question is therefore whether GEM-derived linear programmes converge
reliably on GPUs while reproducing CPU-solver results. GEM-derived
linear programmes are typically degenerate and may be ill-conditioned
when coefficient and bound magnitudes span several orders of magnitude,
so numerical error risks altering feasibility and biological interpretation.}

\textcolor{black}{We present gpuGEM, an open-source Python library
providing an interface between GEMs and cuOpt. Its central contribution
is the systematic identification of size-adaptive solver settings
spanning precision, tolerances, presolve, and stability modes, validated
against CPU reference solvers by objective agreement, primal feasibility,
and bound satisfaction. We report where GPU acceleration provides
an advantage, where CPU solvers remain faster, and which solver settings
are required to maintain feasible solutions.}

\section{Implementation and benchmarking}

\subsection{Library design and solver interface}

gpuGEM is a Python library built on NumPy and SciPy sparse arrays
that formulates FBA as a linear programme and dispatches it to the
PDLP solver in NVIDIA cuOpt \citep{applegate_practical_2021,nvidia_cuopt_2025}.
The core entry point solves
\[
\max_{v}\;c^{\top}v\quad\text{s.t.}\quad Sv=b,\;\;d_{\mathrm{lb}}\le Cv\le d_{\mathrm{ub}},\;\;l\le v\le u,
\]
where $S$ is the stoichiometric matrix, $b$ the right-hand side,
$l$ and $u$ are the lower and upper flux bounds, $c$ the objective
vector, and $C$ is the optional coupling-constraint matrix with lower
and upper bounds $d_{\mathrm{lb}}$ and $d_{\mathrm{ub}}$. The problem
is assembled in the cuOpt data model, solved on the GPU, and returned
as a result object containing the objective value, flux vector, termination
status, feasibility residuals, and solve time. A COBRApy \citep{ebrahim_cobrapy_2013}
adapter converts a model to this representation in one call, while
COBRA Toolbox \citep{heirendt_creation_2019} models can be exported
to the same arrays.

\subsection{Validat\textcolor{black}{ed size-adaptive default settings}}

\textcolor{black}{Default cuOpt settings failed to converge or exceeded
the predefined feasibility criterion on several benchmark models.
gpuGEM therefore applies size-adaptive settings at a threshold of
$10^{5}$ reactions. Both regimes solve in double precision. Smaller
models set cuOpt's double-precision mode explicitly and use $10^{-8}$
primal and dual tolerances, per-constraint convergence and default
presolve. Larger models leave precision at cuOpt's default, which
also resolves to double, and use PaPILO presolve with default $10^{-4}$
tolerances. Settings were selected through systematic parameter sweeps
(Supplementary Methods, gpuGEM / cuOpt configuration; Table S4).}

\subsection{\textcolor{black}{Benchmarking}}

\textcolor{black}{We benchmarked ten representative models spanning
95 to 1.18 million variables: E. coli core, iML1515, Harvey, Harvetta
\citep{thiele_personalized_2020} and six personalised microbiome-whole-body
models (PMM1 to PMM6). All ten models are plotted in Fig. 1a to 1c.
Benchmarks used an Intel Core i9-10900X CPU, 128 GB RAM and NVIDIA
RTX A4500 GPU. gpuGEM used cuOpt 26.6.0, with HiGHS 1.14.0 \citep{huangfu_parallelizing_2018}
and Gurobi 13.0.2 \citep{gurobi_optimizer_2025} as CPU references.
Gurobi barrier was used as the commercial reference, and reported
times are medians of three repeats of single cold solves; full protocols
are provided in Supplementary Methods (Hardware and software environment,
Validation protocol, Timing protocol).}

\section{\textcolor{black}{Results}}

\subsection{\textcolor{black}{gpuGEM reproduces CPU results across four orders
of magnitude}}

\textcolor{black}{Across the ten benchmark models, gpuGEM returned
an optimal, feasible solution for every model. On eight of these ten
models (Harvey, Harvetta and the six personalised microbiome models)
the objective reaction is pinned by its own bounds \citep{thiele_personalized_2020},
so every solver that returns a feasible point reports the same objective
value; agreement there is a consistency check rather than a measure
of accuracy. On the two models whose objective is genuinely optimised,
$E.\ coli$ core and iML1515, gpuGEM agreed with Gurobi and HiGHS
to within $2\times10^{-8}$ relative difference. The accuracy claim
on the remaining models therefore rests on the feasibility residuals,
which remained below the predefined $10^{-4}$ acceptance criterion
(Supplementary Tables\,S2 and S3). All three solvers therefore satisfied
the validation criteria, allowing performance comparisons to focus
on runtime.}

\subsection{\textcolor{black}{gpuGEM outperforms HiGHS on large models}}

\textcolor{black}{Against the open-source CPU solver HiGHS, gpuGEM
was faster on every model above $10^{4}$ variables. On Harvey the
speed-up was 19.7$\times$ against HiGHS's default dual simplex and
8.0$\times$ against whichever of its dual simplex and interior-point
methods is faster on that model; on Harvetta it was 19.3$\times$
and 10.1$\times$, and on the personalised microbiome models 6.1$\times$
to 94$\times$ and 1.1$\times$ to 94$\times$ (Fig. \ref{fig:envelope}a,b).
HiGHS interior point failed on PMM2 to PMM4 (Supplementary Table S8).}

\subsection{gpuGEM has a bounded performance advantage over Gurobi}

\begin{figure}[H]
\centering
\begin{centering}
\includegraphics[width=1\textwidth]{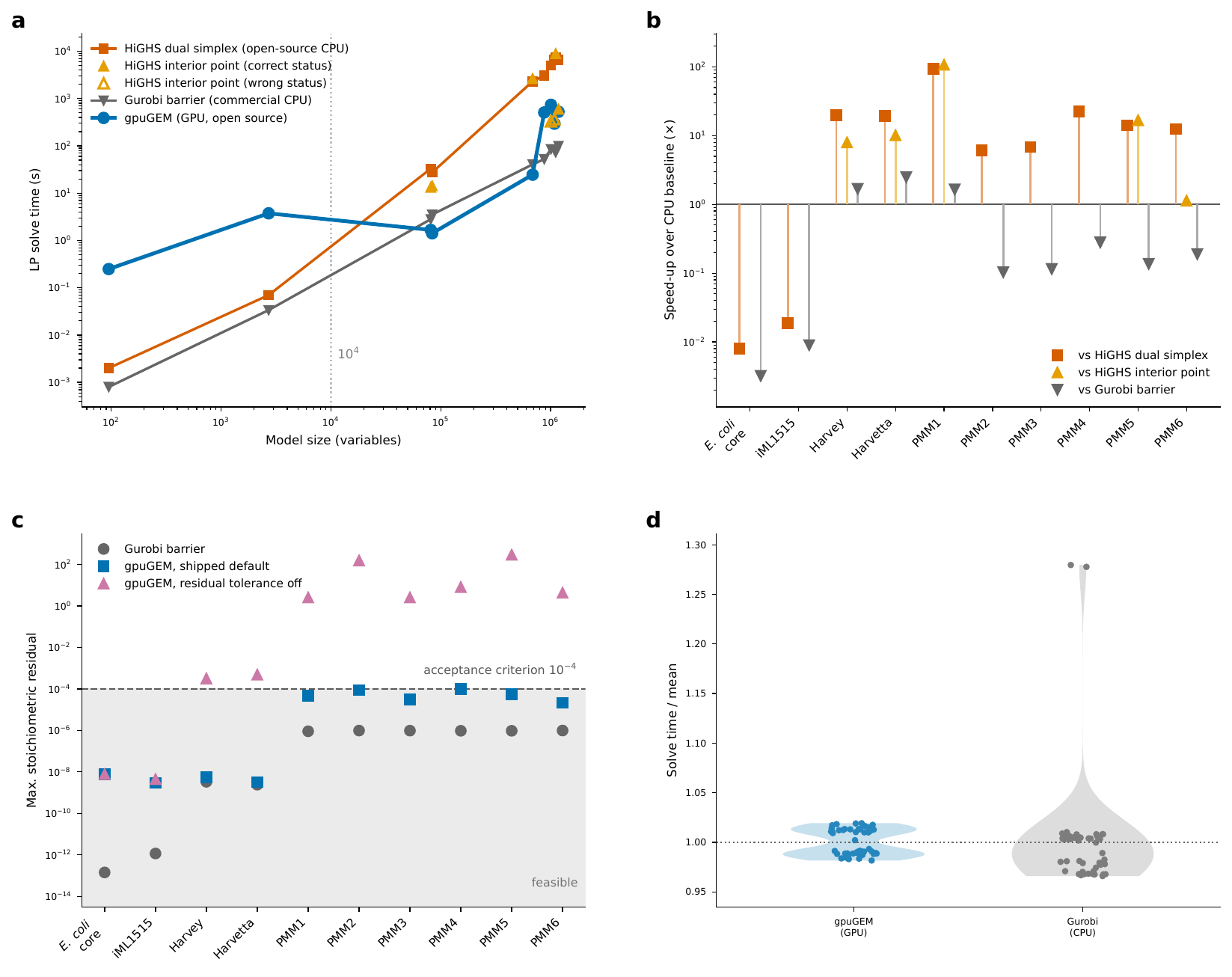}
\par\end{centering}
\caption{\textcolor{black}{gpuGEM performance envelope, numerical correctness
and runtime stability. (a) Median solve time against LP size for gpuGEM,
HiGHS and Gurobi barrier on log-log axes. The dotted line marks $10^{4}$
variables; open triangles mark HiGHS interior-point runs that returned
an incorrect status. (b) Speed-up relative to HiGHS and Gurobi; the
horizontal line marks equal runtime, with values above it indicating
that gpuGEM is faster. (c) Maximum stoichiometric residual across
all ten models for Gurobi, gpuGEM with its shipped defaults (per-constraint
residual checking on), and gpuGEM with cuOpt's default of per-constraint
residual checking off. The shaded region denotes the $10^{-4}$ feasibility
acceptance criterion. (d) Normalised solve time across a 50-objective
panel on PMM1, calculated relative to each solver\textquoteright s
mean. Solve times are medians of three repeats; per-model values are
given in Supplementary Tables S2, S3, S7 and S8.}}

\label{fig:envelope}
\end{figure}

\textcolor{black}{Against commercial Gurobi, gpuGEM showed a bounded
performance advantage rather than one that increased with model size
(Fig.\, \ref{fig:envelope}b). gpuGEM was faster on three of ten
benchmark models: the Harvey and Harvetta whole-body models, by 1.7$\times$
and 2.5$\times$, respectively, and the smallest personalised microbiome
model, with 686,000 variables, by 1.6$\times$. This defines an observed
performance window from approximately $8\times10^{4}$ to $6.9\times10^{5}$
variables. Below this range, Gurobi was more than two orders of magnitude
faster on }\textcolor{black}{\emph{E. coli}}\textcolor{black}{{} core
and iML1515. This difference coincided with PDLP requiring 820$\times$
and 4,785$\times$ more iterations than Gurobi barrier, respectively.
Above the observed window, Gurobi was three to ten times faster on
the remaining five microbiome models (Supplementary Table S3).}

\subsection{\textcolor{black}{Runtime follows problem geometry through PDLP iteration
count}}

\textcolor{black}{gpuGEM runtime tracked PDLP iteration count, which
first-order theory ties to the geometry of the LP (its sharpness)
rather than to its size \citep{applegate_faster_2023,xiong_computational_2023}.
Across the ten benchmark models, log runtime correlated more strongly
with log iteration count (r = 0.97) than with log model size (r =
0.84), while PMM6 solved 213\,s faster than the smaller PMM3 with
331,800 fewer iterations (Supplementary Figure S4).}

\subsection{Relaxing residual checking sacrifices feasibility}

The largest apparent speed-ups arose from configurations that did
not enforce per-constraint feasibility. Disabling per-constraint residual
checking made the tested microbiome-scale model solve up to 72$\times$
faster, but returned 372,156 mass-balance violations (rows with residual
above the criterion) and a maximum residual of $1.6\times10^{2}$,
exceeding the predefined $10^{-4}$ feasibility criterion by several
orders of magnitude (Fig.\,\ref{fig:envelope}c). The relaxed configuration
exceeded this criterion on all five displayed models above $10^{4}$
variables and on neither of the two smaller models. Of seventeen configurations
screened on a microbiome-scale model, the four that satisfied the
validation criteria were no faster than gpuGEM\textquoteright s shipped
default (Supplementary Table\,S4).

\subsection{Runtime is insensitive to the choice of objective on microbiome-scale
models}

\textcolor{black}{To assess sensitivity to biological objective, we
repeated solves across panels of distinct objective functions. On
PMM1, runtime varied less across 50 objectives for gpuGEM than for
Gurobi barrier (CV 1.3\% versus 6.0\%; Fig.\,\ref{fig:envelope}d).
Across all six microbiome models gpuGEM runtime varied by at most
12.9\% across the objective panel (CV 0.1\% to 12.9\%), comparable
to Gurobi barrier (2.8\% to 9.0\%), and gpuGEM gave the lower coefficient
of variation on four of the six. The panel median was within 3\% of
the primary-objective solve time on every microbiome model, so the
timings reported here are representative of the panel rather than
specific to one objective. Extended objective panels across the remaining
models are provided in Supplementary Figure\,S3 and Table\,S7.}

\subsection{Application: GPU-accelerated candidate-gene prioritisation}

Candidate-gene prioritisation from patient biomarkers is a useful
setting for accelerated optimisation because the same large whole-body
model may be queried repeatedly as a patient\textquoteright s biomarker
profile is refined or alternatives are considered. To examine this
use case, we integrated gpuGEM into an inherited-metabolic-disease
candidate-gene prioritisation workflow based on the Harvey whole-body
model and translated from a COBRA Toolbox implementation \citep{heirendt_creation_2019}.
The workflow constrains the model according to measured blood-biomarker
directions and solves two optimisation problems per query: a linear
programme maximising the biomarker-demand objective, followed by a
minimum-norm quadratic programme selecting a unique solution on the
optimal face. Knockout signatures are precomputed using one aggregated
linear programme per disease gene, reducing the required solve count
from hundreds of thousands to hundreds. A single query required 41\,s
using the CPU workflow, including 24.2\,s for the linear-programming
step. Replacing this step with gpuGEM reduced its runtime to 2.7\,s
and total query time to 21\,s, while the quadratic programme remained
on the CPU. Only the linear programme is offloaded here, so the quadratic
programme becomes the dominant cost once the linear step is accelerated,
and the end-to-end speed-up is correspondingly smaller than the speed-up
of the linear step alone. cuOpt's barrier method does support convex
quadratic programmes; moving this step to the GPU was not attempted
in this work and is left as future work. In all three biomarker queries
the causal gene ranked first under both the Gurobi and the gpuGEM
LP engines. The two rankings were strongly concordant over the full
shared candidate list (Spearman's $\rho$ = 0.970, 0.980 and 0.982),
and the ten top-ranked genes agreed on 9, 9 and 10 of 10 (Supplementary
Table\,S10). This experiment demonstrates workflow integration and
computational acceleration rather than diagnostic performance.

\section{\textcolor{black}{Funding}}

This study was funded by grants from the European Research Council
(ERC) under the European Union\textquoteright s Horizon Europe research
and innovation programme (grant agreement \#101125633) to IT, the
Science Foundation Ireland under Grant number 12/RC/2273P2, and the
Horizon Europe grant Recon4IMD (\#101080997). This work was further
supported by the Swiss State Secretariat for Education, Research and
Innovation, SERI (grant 23.00232), and UK Research and Innovation,
UKRI (grants 10083717 and 10080153).

\section{Conflict of interest}

None declared.

\section{Data availability}

The genome-scale metabolic models used in this study are publicly
available from the sources cited in the article, Zenodo and GitHub.
The software, benchmarking code and data underlying the results are
archived at Zenodo at \url{https://zenodo.org/records/23422813}.

\bibliographystyle{plainnat}
\bibliography{GPU_GEM_cuOpt}

\end{document}